\documentclass[useAMS,usenatbib]{mn2e}
\usepackage{graphicx}

\def\be{\begin{equation}} 
\def\ee{\end{equation}}

\def\HI{\hbox{H~$\scriptstyle\rm I\ $}} 
\def\HII{\hbox{H~$\scriptstyle\rm II\ $}}

\def\gsim{\lower.5ex\hbox{\gtsima}} 
\def\lsim{\lower.5ex\hbox{\ltsima}} \def\gtsima{$\; \buildrel > \over 
\sim \;$} \def\ltsima{$\; \buildrel < \over \sim \;$} \def\prosima{$\; 
\buildrel \propto \over \sim \;$} \def\gsim{\lower.5ex\hbox{\gtsima}} 
\def\lsim{\lower.5ex\hbox{\ltsima}} 
\def\simgt{\lower.5ex\hbox{\gtsima}} 
\def\simlt{\lower.5ex\hbox{\ltsima}} 
\def\simpr{\lower.5ex\hbox{\prosima}} \def\la{\lsim} \def\ga{\gsim}

\def\gtsima{$\; \buildrel > \over \sim \;$} 
\def\ltsima{$\; \buildrel < \over \sim \;$} 
\def\gsim{\lower.5ex\hbox{\gtsima}} 
\def\lsim{\lower.5ex\hbox{\ltsima}} 
\def\simgt{\lower.5ex\hbox{\gtsima}} 
\def\simlt{\lower.5ex\hbox{\ltsima}} 
\def\simpr{\lower.5ex\hbox{\prosima}} 
\def\la{\lsim} 
\def\ga{\gsim}

\def\E3{{\cal E}_{\rm g}^{III}}

\title[LAE at $z=8.6$]{Ancient giants: on the farthest galaxy at $z=8.6$}
\author[Dayal \& Ferrara]{Pratika Dayal$^{1,2}$\thanks{E-mail: dayal@aip.de}, \& Andrea Ferrara$^{3}$\\ 
$^{{1}}$ Leibniz Institute  for Astrophysics, An der Sternwarte 16, Potsdam, Germany, 14482\\ 
$^{2}$ Institute of Computational Cosmology, Department of Physics, University of Durham, Science Laboratories, South Road, Durham DH1 3LE \\ 
$^{3}$ Scuola Normale Superiore, Piazza dei Cavalieri 7, 56126 Pisa, Italy}

\begin{document} 
\date{} 
\pagerange{\pageref{firstpage}--\pageref{lastpage}} \pubyear{} 
\maketitle 

\label{firstpage} 
\begin{abstract}
The observational frontiers for the detection of high-redshift galaxies have recently been pushed to unimaginable distances with the record-holding Lyman Alpha Emitter (LAE) UDFy-38135539 discovered at redshift $z=8.6$. However, the physical nature and the implications of this discovery have yet to be assessed. By selecting galaxies with observed luminosities similar to UDFy-38135539 in state-of-the-art cosmological simulations tuned to reproduce the large scale properties of LAEs, we bracket the physical nature of UDFy-38135539: it has a star formation rate $\sim 2.7-3.7 M_\odot$ yr$^{-1}$, it contains $~ 10^{8.3-8.7} M_\odot$ of stars 50-80 Myr old, with stellar metallicity $\sim 0.03-0.12 Z_\odot$. For any of the simulated galaxies to be visible as a LAE in the observed range, the intergalactic neutral hydrogen fraction at $z=8.6$ must be $\chi_{HI} \leq 0.2$ and extra ionizing radiation from sources clustered around UDFy-38135539 is necessary. Finally, we predict that there is a 70\% (15\%) probability of detecting at least 1 such source from JWST (HST/WFC3) observations in a physical radius $\sim 0.4$ Mpc around UDFy-38135539. 
\end{abstract} 


\section{Introduction} 
\label{intro}
The earliest galaxies which ushered in the era of the cosmic dawn changed the state of the intergalactic medium (IGM) out of which they formed in numerous ways: they polluted it with heavy elements, heated and (re)ionized it, thereby affecting the evolution of all subsequent generations of galaxies. We are now in the golden era for the search for such galaxies, made possible by the use of ingenious selection techniques. The first of these, the standard dropout technique (e.g. Steidel et al. 1996; Giavalisco et al. 2004) relies on broad band filters to detect the Lyman break at 912 \AA\, in the galaxy rest frame. Although galaxy candidates have been detected upto $z \sim 10$ (Bouwens et al. 2009) using this method, it has the drawback that the exact source redshift cannot be determined with complete confidence. In this respect, searches for the Lyman Alpha (Ly$\alpha$) line at 1216 \AA\, (in the galaxy rest frame) using narrow band filters have been far more successful in confirming the detection of high-redshift galaxies; specific spectral signatures including the strength, width and continuum break bluewards of the Ly$\alpha$ line make the detection of LAEs (galaxies showing a strong Ly$\alpha$ line) rather unambiguous. Narrow band searches have enabled the confirmation of hundreds of LAEs in a wide high-redshift range, at  $z\approx 5.7$ (Malhotra et al. 2005; Shimasaku et al. 2006), $z\approx 6.6$ (e.g. Kashikawa set al. 2006) and $z \approx 7$ (e.g. Vanzella et al. 2010). Using the same technique, recently, a LAE, designated UDFy-38135539, has been confirmed at $z=8.6$ (Lehnert et al. 2010), making it the farthest astrophysical object known so far; it has overtaken the redshift record of $z=8.2$ set by the Gamma Ray Burst GRB090423 (Salvaterra et al 2009; Tanvir et al. 2009).

This object has already been observed by a number of groups: Lehnert et al. (2010) have observed both the Ly$\alpha$ and ultraviolet (UV) luminosity for UDFy-38135539, and Finkelstein et al. (2010) have obtained broad band UV and Spitzer data points for this galaxy, designated ID 125 in their work. However, the physical nature of the galaxy and the cosmological implications of its discovery have yet to be assessed. In this work, we use state-of-the-art cosmological simulations coupled to a previously developed LAE model (Dayal, Ferrara \& Gallerani 2008; Dayal et al. 2009; Dayal, Ferrara \& Saro 2010), tuned to reproduce a number of observables of LAEs, to bracket the physical properties of UDFy-38135539. We use the observed Ly$\alpha$ luminosity to get a hint on the ionization state of the IGM at $z \sim 8.6$, calculate how much this galaxy could have contributed to reionization and make predictions for the number of Lyman Break Galaxies (LBGs) that could be detected in its vicinity.

\section{Theoretical model}
\label{model}
We start from the analysis of a $z=8.6$ snapshot of a set of cosmological simulations carried out using the TreePM-SPH code {\small {GADGET-2}} (Springel 2005) with the implementation of chemodynamics as described in Tornatore et al. (2007). The periodic simulation box has a comoving size of $75 h^{-1} {\rm Mpc}$ and initially contains $512^3$ particles each of dark matter (DM) and gas.  The masses of the DM and gas particles are $m_{\rm DM}\simeq 1.7\times 10^8\,h^{-1}{\rm M}_\odot$ and $m_{\rm gas}\simeq 4.1\times 10^7\,h^{-1}{\rm M}_\odot$, respectively. For each ``bona-fide'' galaxy (which has at least 20 bound particles; see Saro et al. 2006 for details) in the simulated volume at $z = 8.6$, we compute the  DM halo/stellar/gas mass, star formation rate (SFR), mass-weighted age, gas/stellar metallicity, and gas temperature. The adopted cosmological model for this work corresponds to the $\Lambda$CDM Universe with $\Omega_{\rm m }=0.26, \Omega_{\Lambda}=0.74,\ \Omega_{\rm b}=0.0413$, $n_s=0.95$, $H_0 = (100 h) = 73$ km s$^{-1}$ Mpc$^{-1}$ and $\sigma_8=0.8$, thus consistent with the 5-year analysis of the WMAP data (Komatsu et al. 2009). Complete details of these simulation runs can be found in Dayal et al. (2009).

\subsection{Intrinsic luminosities and dust}
\label{int_lum_dust}

Star forming galaxies produce their intrinsic Ly$\alpha$ line and UV continuum luminosities ($L_\alpha^{int}$ and $L_c^{int}$, at restframe wavelengths  $\lambda_\alpha =1216$ \AA \ and $\lambda_c = 1700$ \AA , respectively) both via stellar (and nebular) emission, and cooling radiation from collisionally excited neutral hydrogen (\HI) in their interstellar medium (ISM). While the spectral energy distributions (SEDs) are obtained using the population synthesis code {\small STARBURST99} (Leitherer et al. 1999) to calculate the stellar and nebular emission, the latter depends on the temperature distribution in the ISM gas. The interested reader is referred to previous work (Dayal, Ferrara \& Saro 2010) for a comprehensive discussion.

As Ly$\alpha$ and continuum photons are efficiently absorbed by dust grains, we calculate the dust content of each galaxy by considering SNII to be the main dust producers; this is justified by the fact that the typical evolutionary time-scale of evolved stars ($\geq 1$ Gyr) becomes longer than the age of the Universe above $z \gsim 5.7$ (Todini \& Ferrara 2001). We further pose that: (i) ${0.5\, \rm M_\odot}$ of dust is produced per SNII (Todini \& Ferrara 2001; Nozawa et al. 2007), (ii) SNII destroy dust in forward shocks with an efficiency of about 40\% (McKee 1998; Seab \& Shull 1983),  and (iii) a homogeneous mixture of gas and dust is assimilated into further star formation. To calculate the dust optical depth, $\tau$, to UV continuum photons, based on the results obtained at $z\sim 6.6$, we assume that dust is made up of carbonaceous grains and spatially distributed as the stars (Dayal, Ferrara \& Saro 2010). The fraction of continuum photons escaping the galaxy undamped by dust is then $f_c = (1- e^{-\tau})\tau^{-1}$. The analogous quantity for Ly$\alpha$ photons is taken to be $f_\alpha=1$ so as to obtain the {\it maximum possible} Ly$\alpha$ luminosity emerging from the galaxy itself.

\subsection{IGM transmission and source clustering}
\label{igm_tx}
After escaping out of the galactic environment, Ly$\alpha$ photons are further attenuated as they travel through the IGM. Depending on the intergalactic hydrogen ionization state, only a fraction $0 < T_\alpha < 1$ of the Ly$\alpha$ photons emerging out of a galaxy undamped by dust reaches the observer (continuum photons are instead unaffected by the IGM); $T_\alpha$ hence depends upon the IGM ionization state. Since this value is largely unconstrained, we explore 6 different values of the average IGM \HI fraction at $z=8.6$: $\chi_{HI}=0.2,0.3,0.4,0.5,0.6,0.7$. For each value of $\chi_{HI}$, we compute the nominal radius, $R_I$, of the spherical, ionized HII region around each simulated galaxy. In reality, though, due to source clustering (i.e. when the separation between any two galaxies becomes smaller than either of their ionized region radii), multiple galaxies can contribute ionizing photons to the same ionized region,
which will then be characterized by an {\it effective} radius, $R_I^{eff} > R_I$, calculated as follows. Suppose galaxies `A' and `B',  and `B' and `C' have overlapping ionized regions. Then, the effective ionized volume `A' is embedded in is the sum of the ionized volumes of `A', `B' and `C'; the same holds true for both `B' and `C'. Within this ionized volume, the total photoionization rate at distance $r$ from `A', is
\begin{eqnarray*}
\Gamma_{tot}^A(r)  & \simeq & \int_{\nu_L}^\infty \frac{L_{\nu,A}^{em}}{4 \pi r^2} \frac{\sigma_L}{h \nu} \bigg(\frac{\nu_L}{\nu}\bigg)^3  d\nu   \\
& + & \sum_{i=1, i \neq A} ^ N \int_{\nu_L}^\infty \frac{L_{\nu,i}^{em}}{4 \pi r_{iA}^2} \frac{\sigma_L}{h \nu} \bigg(\frac{\nu_L}{\nu}\bigg)^3  d\nu + \Gamma_B , 
\end{eqnarray*}
where the terms on the right hand side represent the photoionization rate from (i) the direct radiation from `A', (ii) the galaxies clustered around `A' and (iii) the ultraviolet background (UVB) from distant galaxies, respectively. The UVB photoionization rate is related to $\chi_{HI}$  by $\Gamma_B = (1-\chi_{HI})^2 \chi_{HI}^{-1} n_H \alpha_B$ where $n_H$ is the mean IGM hydrogen number density and $\alpha_B$ is the case-B hydrogen recombination coefficient. Further, $L_{\nu,A}^{em}= L_\nu^{int} f_{esc}$, is the specific ionizing luminosity emerging from `A' and $f_{esc}=0.02$ is the escape fraction of \HI ionizing photons (Gnedin et al. 2008),  $L_{\nu,i}^{em}$ is the ionizing luminosity emerging from the $i^{th}$ galaxy of the total `N' 
galaxies with which `A' shares an ionized region, $\nu_L$ is the frequency corresponding to the Lyman limit wavelength (912 \AA), $\sigma_L$ is the hydrogen photoionization cross-section and $r_{iA}$ is the distance between galaxies $i$ and `A'. This procedure is carried out for each galaxy in the simulated volume. We then assume photoionization equilibrium to compute $\chi_{HI}$ within the effective ionized region of each galaxy. At the edge of this region, we force $\chi_{HI}$ to attain the assigned global value. We use the complete Voigt profile to calculate the optical depth of Ly$\alpha$ photons along the line of sight to compute $T_\alpha$. The observed Ly$\alpha$ and UV continuum luminosity are then simply $L_\alpha = L_\alpha^{int} f_\alpha T_\alpha = L_\alpha^{int} T_\alpha$ (the latter equality descends from our maximal assumption $f_\alpha=1$), and $L_c = L_c^{int} f_c$, respectively. 

\begin{figure} 
\center{\includegraphics[scale=0.48]{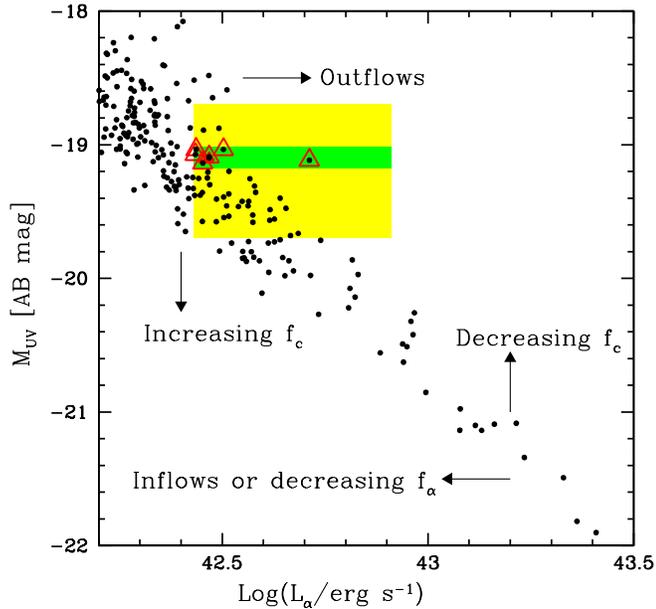}} 
\caption{The UV magnitude plotted as a function of the observed Ly$\alpha$ luminosity for LAEs at $z=8.6$. The yellow shaded area delimitates the region of the parameters deduced from the observation of UDFy-38135539 by Lehnert et al. (2010), the green shaded area shows the UV magnitude range measured for the same galaxy by Finkelstein et al. (2010) and points represent galaxies identified as LAEs in our simulation; the 7 LAE$_{UDF}$ for $\chi_{HI}=0.2$ are marked by triangles. Arrows show the effects of different physical processes on the observed luminosity  (see text for details). }
\label{dist_clus} 
\end{figure} 

As a final step, we select galaxies that would be observable as LAEs according to the canonical criterion, $L_\alpha\geq 10^{42.2} \, {\rm erg \,  s^{-1}}$ {\it and} observed equivalent width $EW \geq 20$ \AA. Among these, we further isolate those that fall within the $L_\alpha$ and continuum magnitude $M_{UV}$ range observed by Lehnert et al. (2010) and Finkelstein et al. (2010): using SINFONI, the former find $L_\alpha = (2.7-8.3) \times 10^{42} \,{\rm erg \, s^{-1}}$, $M_{UV} = -19.6$ to $-18.6$, while Finkelstein et al. (2010) find a tighter bound of $M_{UV} = -19.2$ to $-19.0$. To summarize, all simulated galaxies that have $L_\alpha = (2.7-8.3) \times 10^{42} \,{\rm erg \, s^{-1}}$ and $M_{UV} = -19.2$ to $-19.0$ and designated as LAE$_{UDF}$ in this work. 

\begin{figure} 
\center{\includegraphics[scale=0.48]{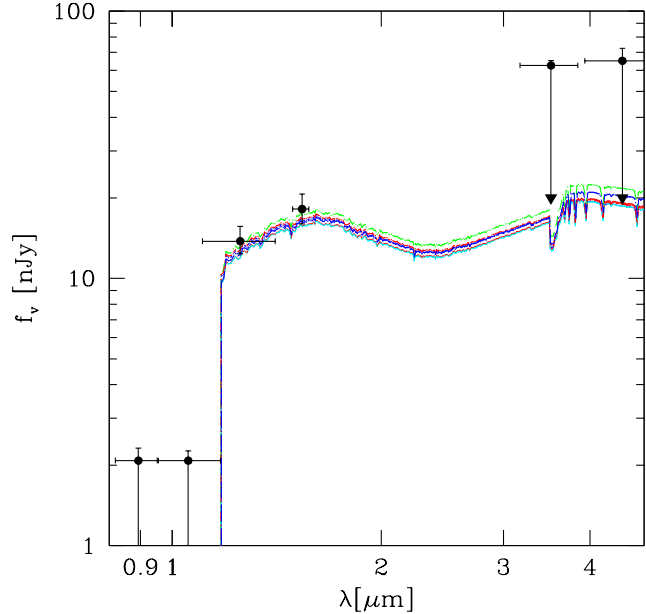}} 
\caption{Comparison of the theoretical SEDs (lines) for the 7 LAE$_{UDF}$ compared to the data (points) collected by Finkelstein et al. (2010); the points at $3.6$ and $4.5\mu$m have been obtained by Spitzer observations and downward pointing arrows show the $1\sigma$ upper limits.}
\label{dist_clus} 
\end{figure}

\section{Results}
\label{results}
We now discuss the main results obtained from the above mentioned computations. We find that, without requiring any tuning of the model parameters, {\it no} LAE$_{UDF}$ are found in the simulated volume for $\chi_{HI} > 0.2$. For $\chi_{HI} =0.2$, we find a total of 215 LAEs in the simulated volume, 7 of which are identified as LAE$_{UDF}$ as shown in Fig. 1. From now on, LAE$_{UDF}$ refer to the LAEs in the combined observed Ly$\alpha$ and UV luminosity range of Lehnert et al. (2010) and Finkelstein et al. (2010) for $\chi_{HI}=0.2$.

We digress here to discuss the two main ingredients whose combined effects can change the slope of the $L_\alpha-M_{UV}$ relation shown in Fig. 1. The first of these concerns the dust attenuation: for any given LAE, an increase (decrease) in $f_c$ leads to the galaxy becoming brighter (fainter) in the UV, shifting the position of the galaxy vertically upward (downward) on the plot; a decrease in the value of $f_\alpha$ (recall that we have used $f_\alpha=1$ in our model) due to dust attenuation of Ly$\alpha$ photons leads to a corresponding decrease in $L_\alpha$, moving the points horizontally leftward on the plot. The second effect is the change in $T_\alpha$ due to peculiar velocities:  inflows (outflows) of neutral gas into (from) the galaxy impart an extra blueshift (redshift) to the Ly$\alpha$ photons, leading to a decrease (increase) in $f_\alpha$ (Santos 2004; Verhamme et al. 2006; Dijkstra \& Wyithe 2010; Dayal, Maselli \& Ferrara 2011) moving the relation horizontally towards the left (right). 

Further constraints on the LAE$_{UDF}$ come from the SEDs which we have obtained from {\small {STARBURST99}} including dust attenuation ($f_c$ ranges between 0.24-0.3 for the LAE$_{UDF}$) using the supernova extinction curve (Bianchi \& Schneider 2007). As shown in Fig. 2, the model SEDs are in excellent agreement with the observed data points, including the two from Spitzer at $3.6$ and $4.5\mu$m obtained by Finkelstein et al. (2010). The SED is a crucial constraint on the physical properties of galaxies that could be identified as LAE$_{UDF}$: although for values of $\chi_{HI}<0.2$ (for the given dust model) a larger number of galaxies would fulfill the LAE$_{UDF}$ selection criterion, their physical properties cannot vary too much without the SEDs becoming inconsistent with the observed one. 

\begin{figure*} 
\center{\includegraphics[scale=0.8]{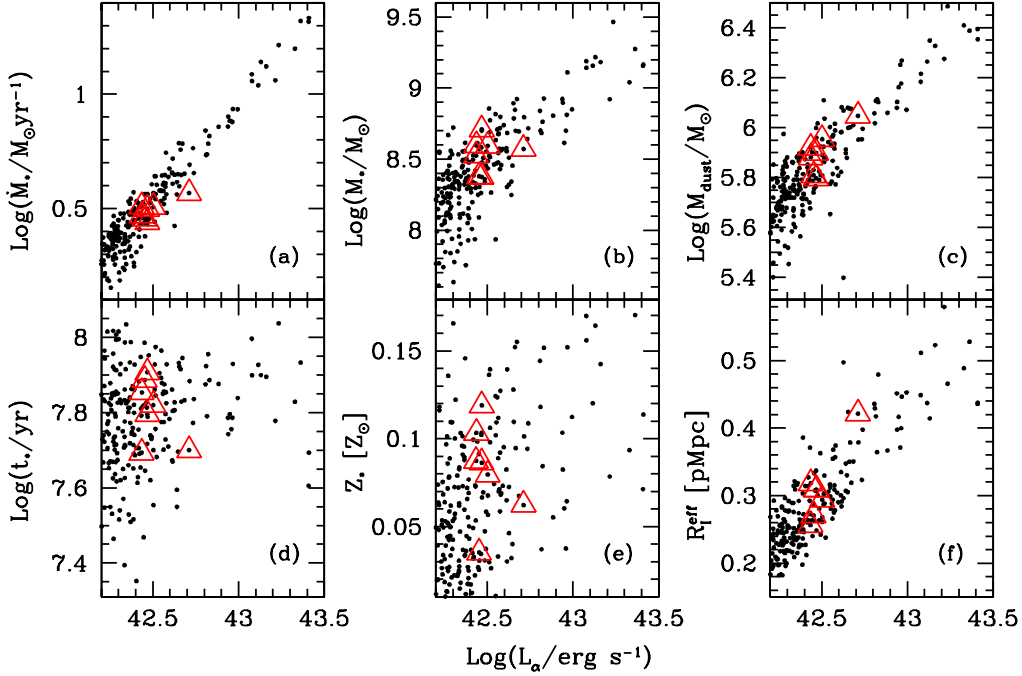}} 
\caption{Summary of the physical properties of LAEs at $z=8.6$.  In each Panel, points represent galaxies identified as LAEs in our simulation; the 7 LAE$_{UDF}$ for $\chi_{HI}=0.2$ (see text for details) are shown by colored symbols. For each LAE, as a function of the observed Ly$\alpha$ luminosity, the Panels represent: (a) the SFR; (b) the stellar mass; (c) the dust mass; (e) the mass-weighted stellar age; (f) the mass-weighted stellar metallicity, and (g) the effective ionized region radius around each LAE.}
\label{phy} 
\end{figure*} 

As the 7 LAE$_{UDF}$ galaxies closely resemble the observed properties of UDFy-38135539, it is reasonable to conclude that their physical properties should also match those of that object, which are now discussed. LAE$_{UDF}$ show SFRs in the range  $2.7-3.7\, M_\odot \, {\rm yr^{-1}}$, as seen from Panel (a) of Fig. 3. These values are quite comparable to the ones inferred by Lehnert et al. (2010) from the Ly$\alpha$ luminosity of UDFy-38135539 ($0.3-2.1\, M_\odot \, {\rm yr^{-1}}$). LAE$_{UDF}$ have stellar masses $\sim 10^{8.3-8.7} \, M_\odot$, (Fig. 3, Panel (b)); this is highly consistent with the stellar mass of $10^{8.6}M_\odot$ found by Finkelstein et al. (2010) from best-fitting the observed SED. In terms of the halo mass, LAE$_{UDF}$ correspond to rare, massive $3\sigma$ mass fluctuations at $z=8.6$ and have halo masses $\sim 10^{10.5} M_\odot$. The LAE$_{UDF}$ have dust masses between $10^{5.7-6} M_\odot$ as seen from Panel (c) of Fig. 3. These lead to values of $f_c =0.24-0.3$ for these galaxies, which translate into $A_V \sim 0.75$ magnitude, which is in excellent agreement with the best fit value of $A_V \sim 0.73$ magnitude derived by Finkelstein et al. (2010) for UDFy-38135539. LAE$_{UDF}$ have ages between 50-80 Myr (see Panel d), in good agreement with the value of 10 to a few 100 Myr inferred observationally by Lehnert et al. (2010). These ages imply that the progenitors of these galaxies started forming as early as $9.2 < z < 9.7$. The stellar metallicity of LAE$_{UDF}$ is 0.03-0.12 $Z_\odot$ (see Panel e), even at $z=8.6$; since heavy elements are the main constituents of grains, this is an additional argument in support of the dusty nature of these galaxies. Indeed dust seems to be required by the measured colors.

For the 7 LAE$_{UDF}$,$R_I^{eff}$ ranges between $0.24-0.42$ physical Mpc (pMpc), as shown in Panel (f), leading to an IGM transmission of $T_\alpha \sim 20$\%. If the effects of clustered sources is neglected, the nominal individual ionized bubble radius is $\sim 0.24$ pMpc, which makes $T_\alpha \sim$ 12\%, rendering these galaxies undetectable in the Ly$\alpha$; the effect of source clustering is {\it crucial} for making these galaxies visible as LAEs. Although the galaxies clustered around the 7 LAE$_{UDF}$ are too faint to be seen in the Ly$\alpha$, nonetheless their UV continuum can be bright enough to be detectable via standard dropout techniques. In Fig. 4 we show the probability (at limiting magnitude $M_{UV} \leq -16$) of such clustered galaxies. Averaged over the 7 LAE$_{UDF}$, there is a 70\% and 15\% probability of JWST and HST/WFC3 (with limits of $M_{UV} <=-16$ and $<=-18$ respectively) finding such a clustered galaxy in a radius $\sim 0.4$ pMpc (or 86 arcsec) at $z=8.6$. 

\begin{figure} 
 \center{\includegraphics[scale=0.48]{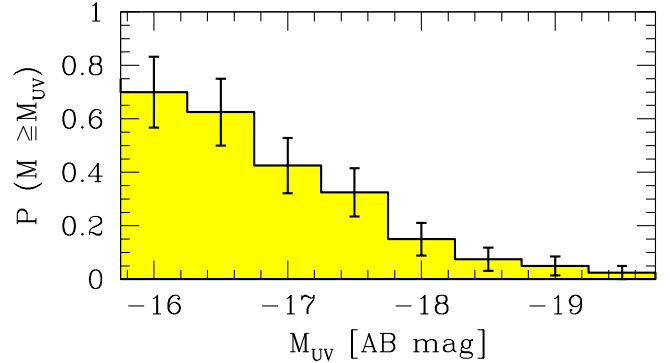}} 
\caption{Average probability of finding a galaxy at least as bright as the magnitude $M_{UV}$ plotted on the x-axis (in bins of 0.5 magnitude), that shares a common ionized region with any of the 7 LAE$_{UDF}$. The error bars show the Poissonian errors.}
\label{dist_clus} 
\end{figure}  

As for the contribution of LAE$_{UDF}$ to reionization, the photon rate density required to balance recombinations, $q_{rec}$, can be expressed as (Madau, Haardt \& Rees 1999)
\begin{equation}
q_{rec} = 10^{51.09} \frac{C}{5} \bigg(\frac{1+z}{9.6}\bigg)^3 \bigg(\frac{\Omega_b h^2}{0.022}\bigg)^2 \, {\rm s^{-1} \, Mpc^{-3}},
\end{equation}
where $C$ is the IGM clumping factor. For the 2\% HI ionizing photon escape fraction used in this work and assuming $C=5$, the LAE$_{UDF}$ have an HI ionizing photon output $\sim 10^{47.2} \, {\rm s^{-1} \, Mpc^{-3}}$; at most, such objects could have contributed $\sim 0.01\%$ of the photons needed to balance recombinations at $z=8.6$.

Finally, we estimate the probability of finding a LAE like UDFy-38135539 in the experimentally sampled volume ($1 \times 10^4 \, {\rm cMpc}^3$). Our simulated volume ($1.08 \times 10^6 \, {\rm cMpc}^3$) is about 100 times larger; therefore we sample the simulated volume by randomly placing the observed volume within it. We find that only 7 of these sub-volumes contain 1 LAE similar to UDFy-38135539 for $\chi_{HI} =0.2$. This translates into a detection probability of only about 7\%, classifying the discovery as a relatively serendipitous one. 

\section{Conclusions and discussion}
\label{conc}
We have constrained the IGM ionization state at $z=8.6$. We find that without requiring any fine-tuning of the model parameters from $z=6.6$ and for the maximum possible Ly$\alpha$ luminosity emerging from the galaxy itself, {\it no} LAE are found in the observed luminosity range of UDFy-38135539 for $\chi_{HI} >0.2$. For $\chi_{HI}=0.2$, we find 7 LAEs (designated LAE$_{UDF}$) whose observed Ly$\alpha$ and UV luminosity fall in the ranges of UDFy-38135539 (Lehnert et al. 2010) and ID 125 (Finkelstein et al. 2010). LAE$_{UDF}$ are observable in the Ly$\alpha$ {\it only} because an overlapping of their \HII regions with those of their nearby galaxies make the effective \HII radius $\sim 0.24-0.42$ pMpc; averaged over the 7 LAE$_{UDF}$, there is a 70\% (15\%) probability of such a clustered source being found by JWST (HST/WFC3) observations.

We have also bracketed the physical properties of UDFy-38135539: the SFR range between $2.7-3.7 \,M_\odot \, {\rm yr^{-1}}$, it has $10^{8.3-8.7} M_\odot$ of stars with a mass weighted age of 50-80 Myr and a mass weighted stellar metallicity between $0.03-0.12 \, Z_\odot$. It is dust enriched with dust masses $\sim 10^{5.7-6} M_\odot$ which translates into $A_V \sim 0.75$ mag and a color excess of $E(B-V) \sim 0.16-0.19$ using the supernova dust extinction curve. 

We add three caveats. The first concerns uncertainties related to dust attenuation of Ly$\alpha$ photons. As has been shown by many works (Neufeld 1991; Hansen \& Oh 2006; Finkelstein et al. 2008; Dayal et al. 2009; Dayal, Ferrara \& Saro 2010; Dayal, Maselli \& Ferrara 2011), the relative escape fraction of Ly$\alpha$ and continuum photons depends on the distribution (smooth/clumpy) of dust in the ISM of high-$z$ galaxies. Lacking data at $z \sim 8$ to support either of the two possibilities, here we have used the maximum value of $f_\alpha=1$. Secondly, an increase (decrease) in the escape fraction of \HI ionizing photons ($f_{esc}$) would decrease (increase) the intrinsic Ly$\alpha$ luminosity, while increasing (decreasing) the sizes of the \HII regions built by each galaxy, thereby affecting the observed Ly$\alpha$ luminosity. Thirdly, outflows might allow Ly$\alpha$ photon transmission also from a substantially more neutral IGM (Dijkstra et al. 2011) than the upper limit of $\chi_{HI} >0.2$ found here. Although interesting, such result has been obtained under the highly idealized conditions of a smooth, spherically symmetric expanding shell of HI gas, whose radius and column density have been tuned to suitable values. As it is very likely that shell fragmentation occurs as a result of Raleigh-Taylor and Kelvin-Helmholtz instabilities arising from the complex gas velocity field around the galaxy (e.g. see Fig. 2 of Fangano et al. 2007), such suggestion needs to await more detailed investigations.

\section{Acknowledgements}
PD acknowledges a Fellowship funded through the Marie Curie Early Stage Training project MEST-CT-2005-021074. PD thanks SISSA for their generous allocation of cluster time and the Scuola Normale Superiore for hospitality. We thank R. Salvaterra and the referee for insightful comments and M. Dijkstra for poitning out an error in an earlier version of the manuscript.

\label{lastpage} 
\end{document}